\documentclass[12pt]{article}
\usepackage{amssymb,amsmath,epsfig}
\allowdisplaybreaks

\begin{document}

\title{\bf Particle Dynamics Near Kerr-MOG Black Hole}

\author{M. Sharif \thanks{msharif.math@pu.edu.pk}~ and
Misbah Shahzadi \thanks{misbahshahzadi51@gmail.com}\\
Department of Mathematics, University of the Punjab,\\
Quaid-e-Azam Campus, Lahore-54590, Pakistan.\\}
\date{}
\maketitle
\begin{abstract}
This paper explores the dynamics of both neutral as well as charged
particles orbiting near a rotating black hole in
scalar-tensor-vector gravity. We study the conditions for the
particle to escape at the innermost stable circular orbit. We
investigate stability of orbits through effective potential and
Lyapunov exponent in the presence of magnetic field. The effective
force acting on particle is also discussed. We also study the center
of mass energy of particle collision near the horizon of this black
hole. Finally, we compare our results with the particle motion
around Schwarzschild, Kerr and Schwarzschild-MOG black holes. It is
concluded that the external magnetic field, spin parameter as well
as dimensionless parameter of the theory have strong effects on
particle dynamics in modified gravity.
\end{abstract}

\section{Introduction}

Mysteries of the universe have always been an interesting topic for
the physicists. Many cosmological observations indicate that the
universe is facing accelerated expansion which is believed due to
the existence of mysterious form of energy named as dark energy. The
matter which cannot be seen directly but can be observed by its
gravitational effects on visible matter is known as dark matter
(DM). This does not interact with electromagnetic force and light.
Modified gravity theories help to uncover the enigmatic nature of
dark energy and DM. These theories are constructed by modifying the
matter or gravitational part of the Einstein-Hilbert action.

Modified theories with additional fields (scalar or vector) like
scalar-tensor theories are the generalization of tensor theory
(general relativity). In Brans-Dicke theory (example of
scalar-tensor theories), gravitational field is obtained by tensor
field $R$ and massless scalar field $\phi$. The vector-tensor
theories are formulated by adding a dynamical vector field coupled
to gravity in the Einstein-Hilbert action. Moffat \cite{2}
formulated a modified theory of gravity (MOG) termed as
scalar-tensor-vector gravity (STVG) which acts as an alternative of
DM. This theory introduces new fields in general relativity which
makes the gravitational field stronger. Its action consists of the
usual Einstein-Hilbert term associated with the metric tensor
$g_{\mu\nu}$, a massive vector field $\phi_{\mu}$ and three scalar
fields which represent the running values of gravitational constant
$G$, coupling constant $\omega$ (determines the coupling strength
between matter and vector field) and the vector field's mass $\mu$
(adjusts the coupling range). The scalar field $G=G_{N}(1+\alpha)$
is the strength of gravitational attraction, where $G_{N}$ is
Newton's gravitational constant and $\alpha$ is a dimensionless
parameter of the theory. The vector field produces a repulsive
gravitational force which is related to a fifth force charge
proportional to mass-energy. This theory helps to explain the solar
system, rotational curves of galaxies, motion of galaxy clusters,
gravitational lensing of galaxy and cluster of galaxies without DM
\cite{3}.

Moffat and Toth \cite{4} studied static spherically symmetric vacuum
solutions for flat FRW model and also discussed the origin of
inertia in STVG theory. Deng et al. \cite{5} discussed the
modifications of STVG and constraint on its parameters. Mishra and
Singh \cite{6} studied the galaxy rotational curves and compared the
form of acceleration law in fourth order gravity with STVG as well
as modified Newtonian dynamics. Moffat and Rahvar \cite{7} used the
weak-field approximation to test the dynamics of cluster of galaxies
and found that this theory is consistent with the observational data
from the solar system to megaparsec scales. Roshan \cite{8}
discussed some cosmological solutions for flat FRW model using
Noether symmetry approach. Sharif and Yousaf \cite{10} used this
approach to find anisotropic exact solutions of locally rotationally
symmetric Bianchi type-I model. Mureika et al. \cite{9} analyzed
thermodynamics of Schwarzschild-MOG as well as Kerr-MOG black hole
(BH) and found a change in entropy area law with the increase of
parameter $\alpha$.

The dynamics of particles (neutral or charged, massive or massless)
around BH is one of the most interesting problems in BH
astrophysics. This plays a key role in understanding the geometrical
structure of spacetime. New observational evidence for BHs provides
new motivations for the investigation of general relativistic
dynamics of particles and electromagnetic fields in the vicinity of
BHs. Astronomical observations over the last decade indicate the
existence of stellar-mass and supermassive BHs in some X-ray binary
systems and in galactic centers.

Hussain and Jamil \cite{12} studied timelike geodesics around
Schwarzschild-MOG BH and found that the stability of orbits
increases due to the presence of vector field in STVG theory.
Pradhan \cite{13} explored circular geodesics near a
Kerr-Newman-Taub-NUT spacetime and found that the energy gain is
maximum for zero NUT parameter and also for maximum spin value.
Babar et al. \cite{14} studied the motion of charged particles in
the vicinity of weakly magnetized naked singularity and explored the
escape velocity of particles orbiting in the inner most stable
circular orbits (ISCOs). Soroushfar et al. \cite{15} discussed the
geodesics around charged rotating BH in $f(R)$ gravity and found
that the shape of an orbit depends on the value of energy, angular
momentum, charge as well as cosmological constant. Sharif and
Iftikhar \cite{16} studied this phenomenon around a higher
dimensional BH and found that higher dimensions have strong effects
on particles motion.

Bardeen \cite{17} investigated characteristics of Kerr BH and its
circular orbits. Aliev and Ozdemir \cite{18} discussed charged
particles motion around rotating BH and found that magnetic field
has strong effect to enlarge the region of stability close to the
event horizon. Frolov and Stojkovic \cite{20} explored particles
motion around five-dimensional rotating BH and found that there does
not exist SCOs in equatorial planes. Aliev and Gumrukcuoglu
\cite{19} studied charged rotating BHs on 3-brane and found that
negative tidal charge increases the horizon radius as well as radii
of photon orbit. Shiose et al. \cite{21} investigated charged
particles motion near a weakly magnetized rotating BH and found that
the radius of ISCO increases due to increase of magnetic field. Amir
et al. \cite{22} studied particle dynamics near a rotating regular
Hayward BH and showed that for particle having angular momentum
$L>L_{c}$ ($L_{c}$ is critical angular momentum), geodesics never
fall into the BH. However, for $L<L_{c}$, the geodesics always fall
into the BH and for $L=L_{c}$, the geodesics fall into the BH
exactly at the event horizon.

The collision energy of particles in the center of mass frame (an
inertial frame in which center of mass is at rest) that results in
the formation of new particles is known as center of mass energy
(CME). This depends upon the nature of colliding particles (e.g.
charged or neutral), astrophysical object (BH or naked singularity)
and gravitational field around the object. When the particles
collision occurs near the horizon, the particles are blue-shifted
due to infinite energy in CM frame \cite{23}. Harada and Kimura
\cite{24} studied the particles collision in ISCO near Kerr BH and
discussed the CME near the horizon. Sharif and Haider \cite{25}
investigated the CME for Demianski and Plebanski BHs without NUT
parameter and examined the dependence of CME on the spin of BH.
Sultana \cite{26} discussed the collision of particles around a Kerr
like BH in Brans-Dicke theory and found that CME is finite whether
the BH is extremal or not. Armaza et al. \cite{27} studied the
spinning massive particles collision in the background of
Schwarzschild BH and found that the CME increases due to the spin of
BH.

In this paper, we discuss the dynamics of particles near a Kerr-MOG
BH in the absence as well as presence of magnetic field. The paper
is organized as follows. In section \textbf{2}, we introduce
Kerr-MOG BH and equations of motion of a neutral as well as charged
particles. Section \textbf{3} explores the behavior of escape
velocity, effective potential, effective force acting on the
particles and instability of the orbits. In section $\textbf{4}$,
the CME for the colliding particles is discussed. In the last
section, we summarize our results.

\section{Dynamics of Particle}

Here we explore the equations of motion for both neutral as well as
charged particles around a Kerr-MOG BH.

\subsection{Neutral Case}

Kerr-MOG BH is the solution of MOG field equations and is fully
described by the mass $M$, spin angular momentum $J=Ma$ and
parameter $\alpha$. The line element around a Kerr-MOG BH is given
as \cite{28}
\begin{eqnarray}\label{1}
ds^{2}=-\frac{\triangle}{\rho^{2}}\left(dt-a\sin^{2}\theta
d\phi\right)^{2}+\frac{\sin^{2}\theta}{\rho^{2}}\left[\left(r^{2}+a^{2}\right)d\phi-adt\right]^{2}
+\frac{\rho^{2}}{\triangle}dr^{2}+\rho^{2} d\theta^{2},
\end{eqnarray}
where
\begin{equation}\nonumber
\triangle=r^{2}+a^{2}-2GMr+\alpha G_{N}GM^{2},\quad
\rho^{2}=r^{2}+a^{2}\cos^{2}\theta.
\end{equation}
Here, $G$ is taken as the gravitational constant, the free parameter
$\alpha$ determines gravitational field strength and $M$ is the mass
of BH. The metric (\ref{1}) is asymptotically flat, stationary and
axially symmetric around $z$-axis. The Killing vectors corresponding
to these symmetries are
\begin{equation}\nonumber
\xi^{\sigma}_{(t)}\partial_{\sigma}=\partial_{t},\quad
\xi^{\sigma}_{(\phi)}\partial_{\sigma}=\partial_{\phi},
\end{equation}
where $\xi^{\sigma}_{(t)}=(1,0,0,0)$ and
$\xi^{\sigma}_{(\phi)}=(0,0,0,1)$. When $\alpha=0$, the metric
(\ref{1}) reduces to the Kerr metric, further $a=\alpha=0$ leads to
the Schwarzschild metric and for $a=0$, we obtain Schwarzschild-MOG
metric. The Kerr-MOG metric $(1)$ is singular if $\rho$ or $\Delta$
vanishes. The curvature and coordinate singularities correspond to
$\rho=0$ and $\Delta=0$, respectively. The horizons of (\ref{1}) can
be obtained by $\Delta=0$ as
\begin{equation}\nonumber
r_{\pm}=GM\pm\sqrt{G^{2}M^{2}-a^{2}-\alpha G_{N}GM^{2}},
\end{equation}
here $\pm$ sign correspond to the event and Cauchy horizons,
respectively. The ergosphere can be obtained by solving $g_{tt}=0$
\begin{equation}\nonumber
r_{es}=GM\pm\sqrt{G^{2}M^{2}-\alpha G_{N}GM^{2}}.
\end{equation}
For $\theta=0, \pi$, both ergosphere and event horizon coincide. The
extremal condition is located at $G^{2}M^{2}=a^{2}+\alpha
G_{N}GM^{2}$. The corresponding angular velocity is
\begin{equation}\nonumber
\Omega_{H}=\frac{a}{(r_{+}^{2}+a^{2})}=\frac{a}{2G^{2}M^{2}-\alpha
G_{N}GM^{2}+2GM\sqrt{G^{2}M^{2}-a^{2}-\alpha G_{N}GM^{2}}}.
\end{equation}
We consider the equatorial plane to find the motion of test
particle, i.e, $\theta=\frac{\pi}{2}$,\quad $\dot{\theta}=0$.

The motion of a neutral particle can be illustrated by the
Lagrangian
\begin{equation}\nonumber
\mathcal{L}=\frac{1}{2}g_{\sigma\eta}\dot{x}^{\sigma}\dot{x}^{\eta},
\end{equation}
where $\dot{x}^\sigma=u^\sigma=\frac{dx^\sigma}{d\tau}$ is the four
velocity of particle and $\tau$ is the proper time. For Kerr-MOG BH,
the Lagrangian becomes
\begin{eqnarray}\nonumber
2\mathcal{L}&=&-\left(\frac{r^{2}-2GMr+\alpha
G_{N}GM^{2}}{r^{2}}\right)\dot{t}^{2}-2a\left(\frac{2GMr-\alpha
G_{N}GM^{2}}{r^{2}}\right)\dot{t}\dot{\phi}\\\label{2}&+&\frac
{r^{2}}{\Delta}\dot{r}^{2}+\frac{\left[r^{2}(r^{2}+a^{2})+a^{2}\left(2GMr-\alpha
G_{N}GM^{2}\right)\right]}{r^{2}}\dot{\phi}^{2}.
\end{eqnarray}
It is clear from Eq.(\ref{2}) that $t$ and $\phi$ are cyclic
coordinates. Corresponding to these cyclic coordinates there are two
constants of motion, i.e., total energy $E$ and azimuthal angular
momentum $L_{z}$ which are conserved along geodesics. The
generalized momenta are
\begin{eqnarray}\nonumber
-p_{t}&=&-\left(\frac{r^{2}-2GMr+\alpha
G_{N}GM^{2}}{r^{2}}\right)\dot{t}-a\left(\frac{2GMr-\alpha
G_{N}GM^{2}}{r^{2}}\right)\dot{\phi}\\\label{3}&=&E,\\\nonumber
p_{\phi}&=&-a\left(\frac{2GMr-\alpha
G_{N}GM^{2}}{r^{2}}\right)\dot{t}\\\label{4}&+&\frac{\left[r^{2}(r^{2}
+a^{2})+a^{2}\left(2GMr-\alpha
G_{N}GM^{2}\right)\right]}{r^{2}}\dot{\phi}=L_{z},\\\nonumber
p_{r}&=&\frac{r^{2}}{\Delta}\dot{r},
\end{eqnarray}
where dot denotes derivative with respect to $\tau$. From
Eqs.(\ref{3}) and  (\ref{4}), we obtain
\begin{eqnarray}\nonumber
\dot{t}&=&\frac{1}{r^{2}\Delta}\left[\left(r^{2}\left(r^{2}+a^{2}\right)+a^{2}\left(2GMr-\alpha
G_{N}GM^{2}\right)\right)E \right.\\\label{5}&-&\left. a
L_{z}\left(2GMr-\alpha G_{N}GM^{2}\right)\right],
\\\label{6} \dot{\phi}&=&\frac{1}{r^{2}\Delta}\left[\left(r^{2}-2GMr+\alpha
G_{N}GM^{2}\right)L_{z}+a\left(2GMr-\alpha
G_{N}GM^{2}\right)E\right].
\end{eqnarray}
The total angular momentum is given as
\begin{equation}\nonumber
L^{2}=\left(r^{2}\dot{\theta}\right)^{2}+\left(r^{2}\dot{\phi}\sin\theta\right)^{2},
\end{equation}
here $\upsilon_{\bot}^{2}\equiv-r\dot{\theta}_{0}$,
$\dot{\theta}_{0}$ is the initial angular velocity of the particle.
Using the value of $\dot{\phi}$ from Eq.(\ref{6}), we have
\begin{eqnarray}\nonumber
L^{2}&=&\frac{\left[\left(r^{2}-2GMr+\alpha
G_{N}GM^{2}\right)L_{z}+a\left(2GMr-\alpha
G_{N}GM^{2}\right)E\right]^{2}}{\left(r^{2}+a^{2}-2GMr+\alpha
G_{N}GM^{2}\right)^{2}}\\\label{a}&+&r^{2}\upsilon_{\perp}^{2}.
\end{eqnarray}

Using the normalization condition,
$g_{\sigma\eta}u^{\sigma}u^{\eta}=-1$, it follows that
\begin{equation}\label{7}
\dot{r}^{2}=E^{2}+\frac{1}{r^{2}}\left(a^{2}E^{2}-L_{z}^{2}\right)+\frac{1}{r^{4}}\left(2GMr-\alpha
G_{N}GM^{2}\right)\left(aE-L_{z}\right)^{2}-\frac{\Delta}{r^{2}}.
\end{equation}
Equations (\ref{5})-(\ref{7}) are useful to discuss various features
of particle motion near (\ref{1}). From Eq.(\ref{7}), we obtain
\begin{equation}\nonumber
\frac{1}{2}\left(E^{2}-1\right)=\frac{1}{2}\dot{r}^{2}+U_{eff}(r,E,L_{z}),
\end{equation}
where
\begin{eqnarray}\nonumber
U_{eff}(r,E,L_{z})&=&\frac{-GM}{r}+\frac{\left[L_{z}^{2}-a^{2}\left(E^{2}-1\right)+\alpha
G_{N}GM^{2}\right]}{2r^{2}}\\\nonumber&-& \frac{GM}{r^{3}}\left(a
E-L_{z}\right)^{2}+\frac{\alpha G_{N}GM^{2}}{2r^{4}}\left(a
E-L_{z}\right)^{2}.
\end{eqnarray}
The maximum and minimum values of effective potential $U_{eff}$
determine the unstable and stable circular orbits, respectively. The
radius of innermost stable circular orbit ($r_{0}$) can be found by
solving $\frac{dU_{eff}}{dr}=0$. We have solved
$\frac{dU_{eff}}{dr}=0$ using Mathematica 8.0 and found three roots
of $r$. We have ignored the two imaginary values of $r$ and the rest
is the radius of ISCO ($r_{0}$). The energy and azimuthal angular
momentum corresponding to $r_{0}$ are
\begin{eqnarray}\label{8}
E_{0}&=&\frac{1}{r_{0}}\left[\frac{r_{0}^{2}-2GMr_{0}+\alpha G_{N} G
M^{2}\mp a \sqrt{G M(r_{0}-M \alpha G_{N})}}
{\sqrt{r_{0}^{2}-3GMr_{0}+2 \alpha G_{N} G M^{2}\mp 2a\sqrt{G
M(r_{0}-M \alpha G_{N})}}}\right], \\\nonumber
L_{z0}&=&\frac{1}{r_{0}^{2}}\left[\mp \sqrt{GM(r_{0}-M\alpha
G_{N})}(a^{2}+r_{0}^{2} \pm 2a\sqrt{GM(r_{0}-M\alpha G_{N})})
\right.\\\nonumber&-&\left.\alpha
G_{N}GM^{2}a\right]\left[r_{0}^{2}-3GMr_{0}+2 \alpha G_{N} G
M^{2}\right.\\\label{9}&\mp&\left. 2a\sqrt{G M (r_{0}-M \alpha
G_{N})}\right]^{\frac{-1}{2}},
\end{eqnarray}
where $r_{0}$ is the radius of ISCO and $\pm$ signs correspond to
the counter-rotating and co-rotating orbits, respectively. For
$\alpha=0$, Eqs.(\ref{8}) and (\ref{9}) reduce to Kerr-BH \cite{29}.

Consider a particle orbiting in ISCO colliding with another particle
which is at rest. After collision, there are three cases (depending
upon the collision process) for the particles motion, i.e, either
captured by BH, or bounded near a BH or escape to infinity. If there
is a small change in energy and angular momentum then the particle's
orbit will slightly be perturbed and particle remains bounded. But
for large change, it may be captured by BH or escape to infinity.
After collision, particle will be in new plane with respect to the
original equatorial plane. Thus the particle would have new energy
and azimuthal angular momentum. For simplicity, we assume that,
after collision, initial radial velocity and azimuthal angular
momentum do not change and particle attains escape velocity
$v_{esc}$ orthogonal to the equatorial plane. The new energy and
angular momentum of particle are
\begin{eqnarray}\nonumber
L^{2}_{new}&=&r_{0}^{2}\upsilon_{\perp}^{2}+\frac{[(r_{0}^{2}-2GMr_{0}+\alpha
G_{N}GM^{2})L_{z0}+a(2GMr_{0}-\alpha
G_{N}GM^{2})E_{0}]^{2}}{(r_{0}^{2}+a^{2}-2GMr_{0}+\alpha
G_{N}GM^{2})^{2}},\\\nonumber
E^{2}_{new}&=&1-\frac{2GM}{r_{0}}+\frac{[L_{z0}^{2}-a^{2}(E_{0}^{2}-1)+\alpha
G_{N}GM^{2}]}{r_{0}^{2}}- \frac{2GM}{r_{0}^{3}}(a
E_{0}-L_{z0})^{2}\\\label{c} &+&\frac{\alpha
G_{N}GM^{2}}{r_{0}^{4}}(a E_{0}-L_{z0})^{2}.
\end{eqnarray}
After collision, particle attains greater energy and angular
momentum as compared to before collision. We observe from the above
equation that $E_{new}\rightarrow1$ as $r_{0}\rightarrow\infty$.
Thus for unbounded motion, particle requires $E_{new}\geq1$ to
escape, whereas particle cannot escape for $E_{new}<1$.

\subsection{Charged Case}

The theoretical and experimental evidences indicate that magnetic
field must be present in the vicinity of BHs. It arises due to
plasma in the surrounding of BH \cite{7r} and plays an important
role in the formation, structure and evolution of planets, stars,
galaxies and possibly the entire universe \cite{8r}. The magnetic
field has strong effects around the event horizon but does not
change the geometry of BH rather the motion of charged particles is
affected \cite{9r}. We assume that a particle has an electric charge
and its motion is affected by magnetic field in the BH exterior.
Since Kerr-MOG BH is non-vacuum, so we follow \cite{30} to calculate
the four-vector potential given as
\begin{eqnarray}\nonumber
A^{\sigma}&=&\left[\frac{-Qr}{r^{2}-2f}+a\breve{B}\left(1+\frac{f_{2}
\sin^{2}\theta}{r^{2}}\right)\right]\xi_{(t)}^{\sigma}
+\left[\frac{\breve{B}}{2}\left(1+\frac{2f_{2}}{r^{2}}\right)\right.\\\nonumber&-&\left.
\frac{Q a}{r\left(r^{2}-2f\right)}\right]\xi_{(\phi)}^{\sigma},
\end{eqnarray}
where $f=f_1r+f_2$ with $f_1,~f_2$ are constants and $\breve{B}$ is
the magnetic field. For simplicity, we take $Q=0$ and also for the
equatorial plane $\theta=\pi/2$. Thus the above equation becomes
\begin{equation}\nonumber
A^{\sigma}=a\breve{B}\left(1+\frac{f_{2}}{r^{2}}\right)\xi_{(t)}^{\sigma}+\frac{\breve{B}}{2}\left(1+
\frac{2f_{2}}{r^{2}}\right)\xi_{(\phi)}^{\sigma}.
\end{equation}
For (\ref{1}), the above equation takes the form
\begin{equation}\nonumber
A^{\sigma}=a\breve{B}\left(1-\frac{\alpha G_{N}G M^{2}}{2
r^{2}}\right)\xi_{(t)}^{\sigma}+\frac{\breve{B}}{2}\left(1-\frac{\alpha
G_{N}G M^{2}}{r^{2}}\right)\xi_{(\phi)}^{\sigma}.
\end{equation}
The magnetic field for an observer having four velocity $u_{\eta}$
is defined as
\begin{equation}\nonumber
\breve{B}^{\sigma}=-\frac{1}{2}e^{\sigma\eta\gamma\delta}F_{\gamma\delta}u_{\eta},
\end{equation}
where
$e^{\sigma\eta\gamma\delta}=\frac{\varepsilon^{\sigma\eta\gamma\delta}}{\sqrt{-g}}$
and $\varepsilon^{\sigma\eta\gamma\delta}$ is the Levi-Civita
symbol, $g=det(g_{\sigma\eta})$ and $\varepsilon_{0123}=1$. The
Maxwell field tensor is
$F_{\sigma\eta}=A_{\eta;\sigma}-A_{\sigma;\eta}$.

The Lagrangian of a particle having electric charge $q$ and mass $m$
is
\begin{equation}\nonumber
\mathcal{L}=\frac{1}{2}g_{\sigma\eta}u^{\sigma}u^{\eta}+\frac{q}{m}A_{\sigma}u^{\sigma}.
\end{equation}
The generalized four momentum is given by
\begin{equation}\nonumber
p_{\sigma}=mu_{\sigma}+qA_{\sigma}.
\end{equation}
In the presence of magnetic field, the constants of motion are
\begin{eqnarray}\nonumber
\dot{t}&=&\frac{1}{r^{3}\triangle}[r(8
a^{3}BG^{2}M^{2}-2aG(2a^{2}B-aE+L)
Mr+a^{2}(-2aB+E)\\\nonumber&\times&r^{2} +(-2aB+E)r^{4})+a\alpha
G_{N}GM^{2}(-aEr+a^{2}B(-4GM+3r)\\\label{10}&+&r(L+B
r(2GM+r))-Br\alpha G_{N}GM^{2})], \\\nonumber
\dot{\phi}&=&\frac{1}{r^{3}\triangle}[r(2aEGMr+a^{2}B(8G^{2}M^{2}-8GMr+r^{2})
-(2GMr-r)r\\\nonumber&\times&(L+Br^{2}))+\alpha
G_{N}GM^{2}(-aEr+a^{2}B(-4GM+3r)+r(L+2B\\\label{11}&\times&G M
r)-Br\alpha G_{N}GM^{2})],
\end{eqnarray}
where $B=\frac{q\breve{B}}{2m}$. Using the normalization condition,
we obtain
\begin{eqnarray}\nonumber
\dot{r}^{2}r^{6}&=&[r(a^{4}B^{2}(4GM-3r)(8G^{2}M^{2}-6GMr-r^{2})-4a^{3}BE
r(-4G^{2}M^{2}\\\nonumber&+&2GMr+r^{2})-4aEr^{2}(GLM+Br^{3})+a^{2}r(-2BL(8G^{2}M^{2}
-8G\\\nonumber&\times&Mr+r^{2})+2B^{2}r^{2}(-6G^{2}M^{2}+4GMr+r^{2})+r(-r+E^{2}(2GM\\\nonumber&+&r)))
+r^{2}(-L^{2}r +(-1+E^{2}-2BL)r^{3}-B^{2}r^{5}+2GM(r^{2}+
(L\\\nonumber&+ &Br^{2})^{2})))-\alpha
G_{N}GM^{2}(16a^{4}B^{2}G^{2}M^{2}-8a^{2}B
G(3a^{2}B-aE+L)\\\nonumber&\times&Mr+((3a^{2}B-aE+L)^{2}
-16a^{2}B^{2}G^{2}M^{2})r^{2}+4BG(2a^{2}B-a\\\nonumber&\times&E+L)Mr^{3}
+(1+2aB(2aB-E))r^{4}+4B^{2}GMr^{5}-B^{2}r^{6}+B
r\\\nonumber&\times&\alpha
G_{N}GM^{2}(a^{2}B(8GM-6r)+2aEr-r(2L+Br(2GM+r))\\\label{12}&+&Br\alpha
G_{N}GM^{2}))],
\end{eqnarray}
which can be written as
\begin{equation}\nonumber
\frac{1}{2}\left(E^{2}-1\right)=\frac{1}{2}\dot{r}^{2}+U_{eff}(r,E,L_{z},B),
\end{equation}
where
\begin{eqnarray}\nonumber
U_{eff}(r,E,L_{z},B)&=&\frac{-1}{2r^{6}}[r(a^{4}B^{2}(4GM-3r)(8G^{2}M^{2}-6GMr
-r^{2})-4a^{3}\\\nonumber&\times& BE
r(-4G^{2}M^{2}+2GMr+r^{2})-4aEr^{2}(GL_{z}M+Br^{3})\\\nonumber&+&a^{2}r(-2BL_{z}(8G^{2}M^{2}
-8GMr+r^{2})+2B^{2}r^{2}(-6G^{2}M^{2}\\\nonumber&+&4GMr+r^{2})
+r(-r+E^{2}(2GM+r)))+r^{2}(-L_{z}^{2}r-2\\\nonumber&\times&BL_{z}r^{3}
-B^{2}r^{5}+2GM(r^{2}+ (L_{z}+Br^{2})^{2})))-\alpha
G_{N}GM^{2}\\\nonumber&\times&(16a^{4}B^{2}G^{2}M^{2}-8a^{2}B
G(3a^{2}B-aE+L_{z})Mr+((3a^{2}B\\\nonumber&-&aE+L_{z})^{2}
-16a^{2}B^{2}G^{2}M^{2})r^{2}+4BG(2a^{2}B-aE+L_{z})\\\nonumber&\times&Mr^{3}
+(1+2aB(2aB-E))r^{4}+4B^{2}GMr^{5}-B^{2}r^{6}+B\\\nonumber&\times&r\alpha
G_{N}GM^{2}(a^{2}B(8GM-6r)+2aEr -r(2L_{z}+B
r(2G\\\label{13}&\times&M+r))+Br\alpha G_{N}GM^{2}))].
\end{eqnarray}
Equations (\ref{10}) to (\ref{13}) are invariant under the
transformations $\phi \rightarrow -\phi,
L_{z}\rightarrow-L_{z},B\rightarrow-B$.

\section{Escape Velocity and Effective Potential}

In this section, we discuss properties of the escape velocity and
effective potential of a particle when $G=1,~M=1,~G_{N}=1,~E=1$.
Figure \textbf{1} shows the escape velocity for a particle moving
around Kerr-MOG BH in the presence of magnetic field. In the upper
panel, the left graph indicates that particles with large angular
momentum have less possibilities to escape as compared to those
having small angular momentum. The right graph shows that the escape
velocity increases with the increase of magnetic field. Particles
attain more energy due to large values of magnetic field and can
easily escape. In the lower panel, the left graph provides
comparison for escape velocity of the Kerr-MOG BH with Schwarzschild
and Kerr BHs. This shows that the Kerr-MOG BH has more $v_{esc}$ as
compared to Schwarzschild and Schwarzschild-MOG BH, also $v_{esc}$
increases with increasing value of parameter $\alpha$.
\begin{figure}\centering
\epsfig{file=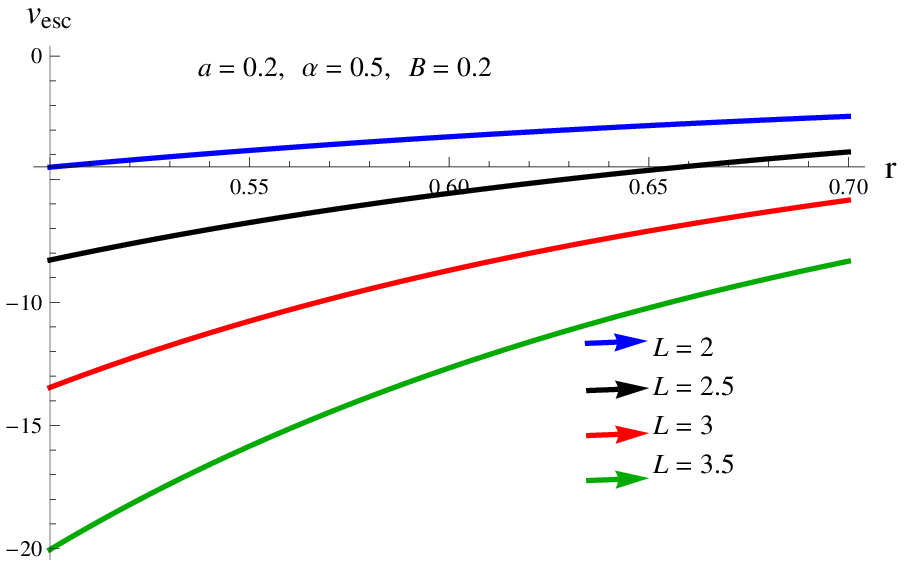,width=.44\linewidth}
\epsfig{file=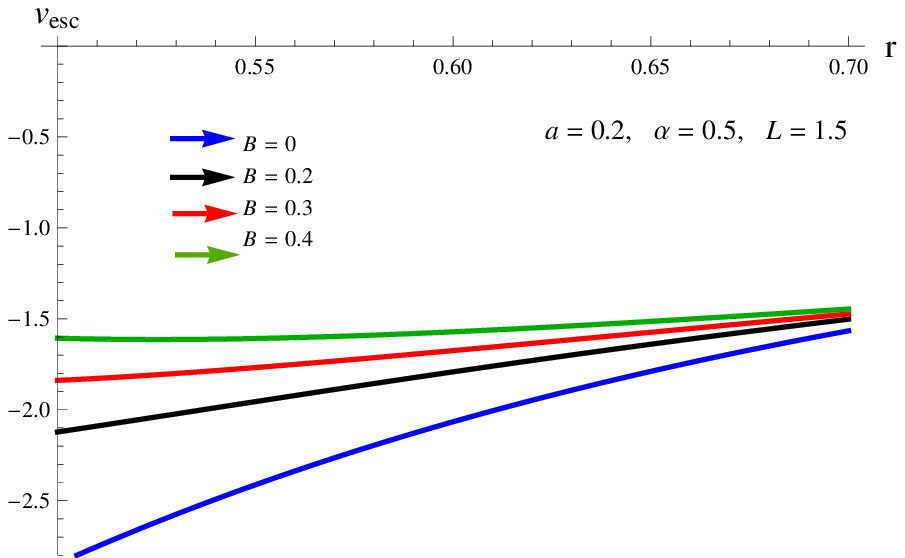,width=.44\linewidth}
\epsfig{file=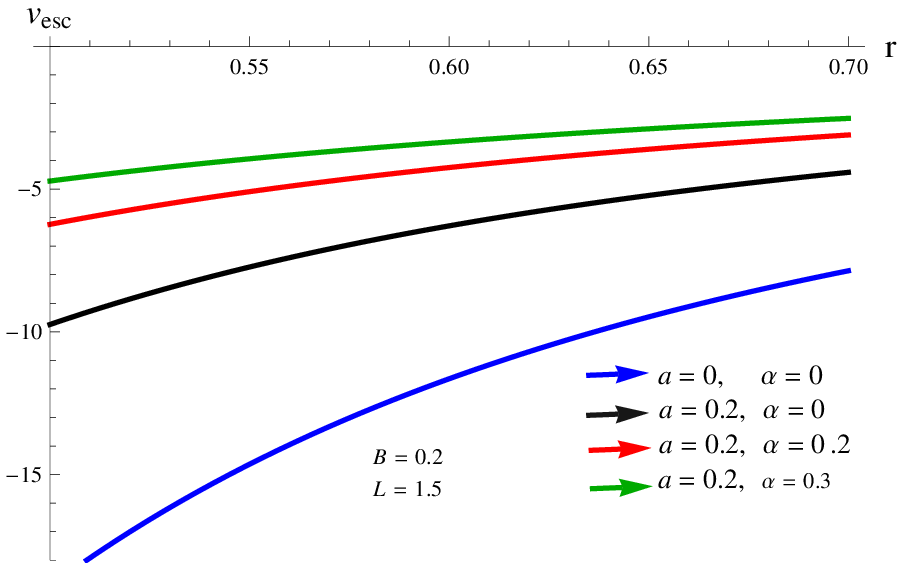,width=.44\linewidth}
\epsfig{file=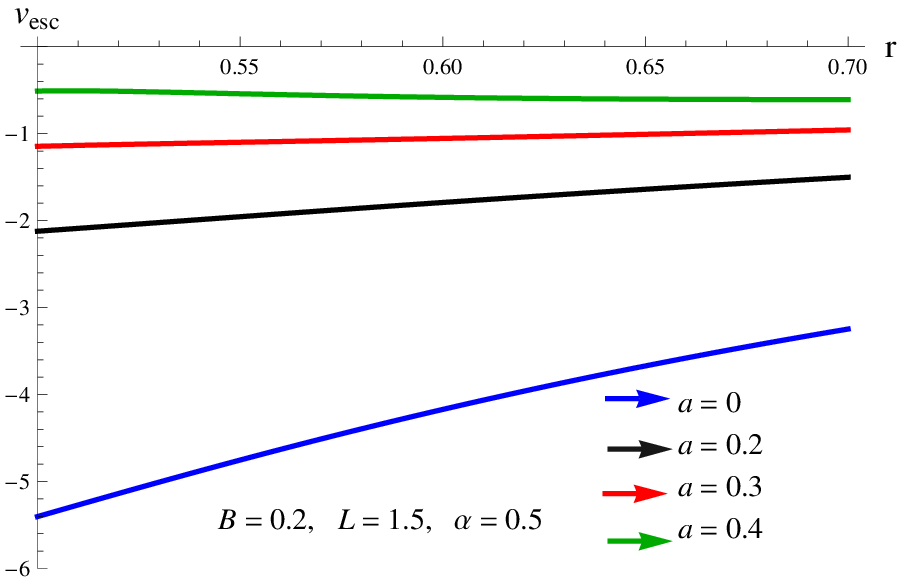,width=.44\linewidth} \caption{Escape velocity
against $r$.}
\end{figure}

The effect of spin on escape velocity is shown in the right graph.
This also provides the comparison of escape velocity of particle
around Schwarzschild-MOG BH with Kerr-MOG BH. We see that with the
increase of spin of BH, particles have more possibilities to escape
as $v_{esc}$ is high for large values of spin parameter $a$. A
rotating BH ($a\neq0$) may provide sufficient amount of energy to
particle due to which it can escape to infinity as compared to
non-rotating BH. It is also noted that particles in the vicinity of
Schwarzschild-MOG BH can escape easily as compared to Kerr-MOG BH.
We conclude that $v_{esc}$ becomes almost constant as particle moves
away from the BH.

Figure \textbf{2} shows the behavior of $U_{eff}$ against $r$. The
stable and unstable circular orbits correspond to minimum and
maximum values of effective potential, respectively. In the upper
panel, the left graph shows that initially orbits are unstable and
then become stable, stability increases with the increase of $L$.
The right graph is plotted for different values of $B$. We observe
that the orbits are initially stable, then becomes unstable and
stability decreases for increasing value of $B$. In the lower panel,
the left graph shows that the particle motion becomes more unstable
for high values of $\alpha$. The right graph indicates that circular
orbits for Kerr-MOG BH are more unstable as compared to Kerr and
Schwarzschild BH. The last graph is plotted for different values of
spin parameter $a$ which indicate that the stability of circular
orbits decreases with the increase of $a$. Thus the motion of
particle will be more unstable for large value of $a$. We note that
circular orbits around Schwarzschild-MOG BH are more stable as
compared to Kerr-MOG BH.
\begin{figure}\centering
\epsfig{file=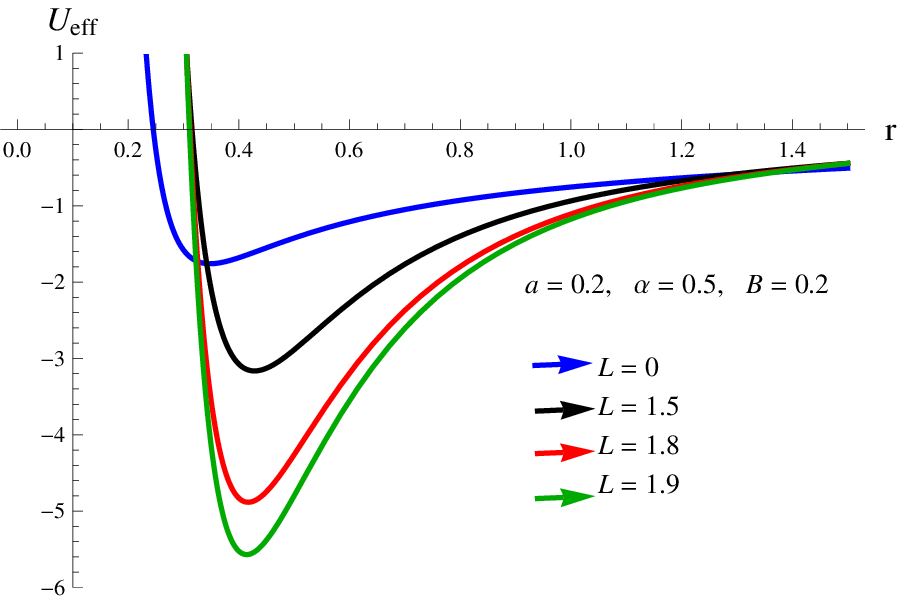,width=.44\linewidth}
\epsfig{file=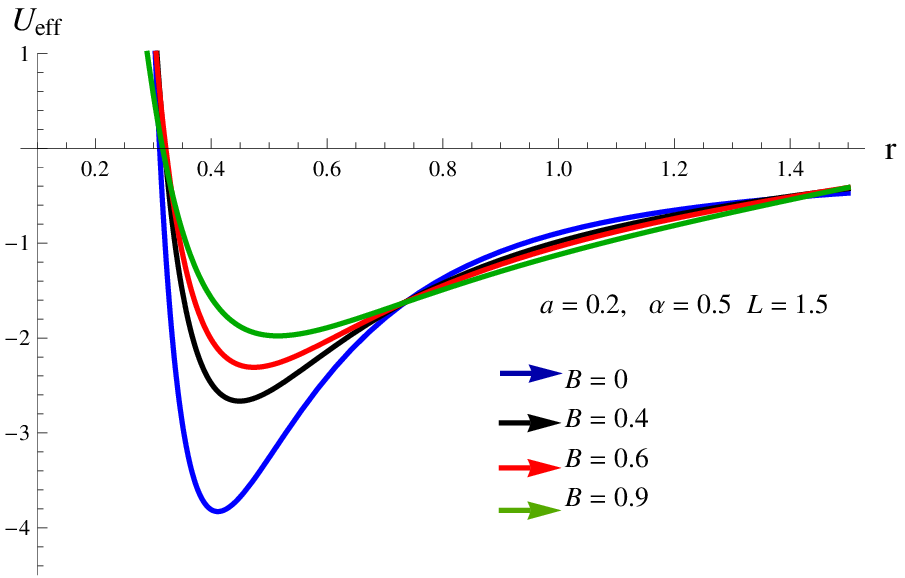,width=.44\linewidth}
\epsfig{file=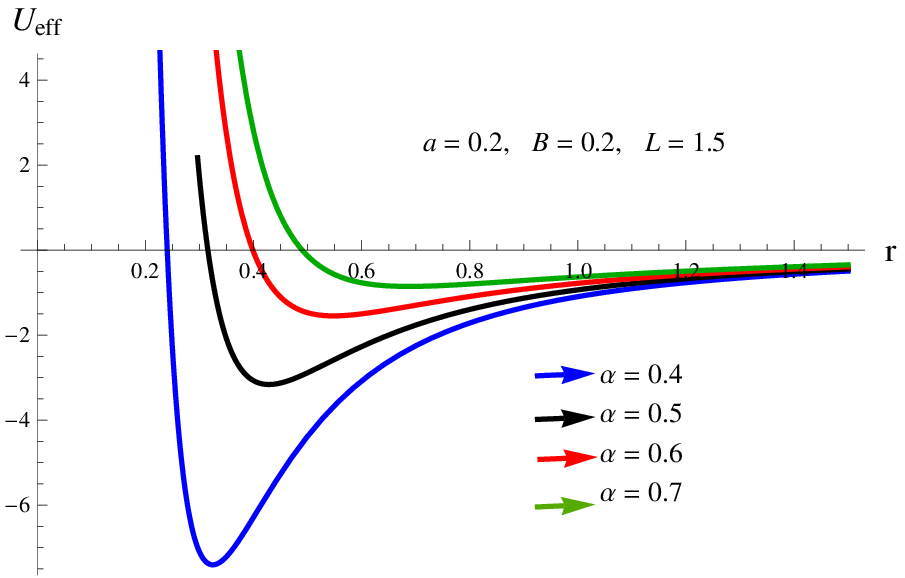,width=.44\linewidth}
\epsfig{file=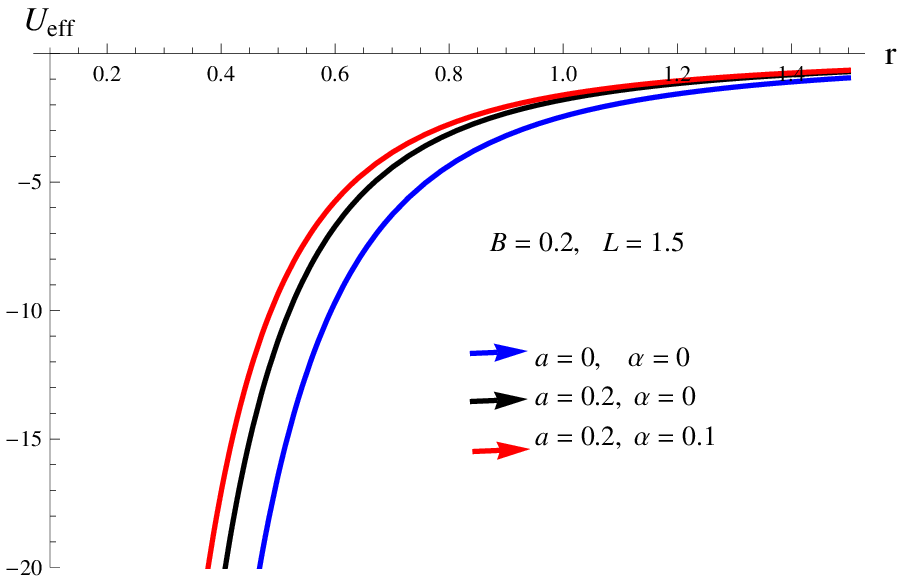,width=.44\linewidth}
\epsfig{file=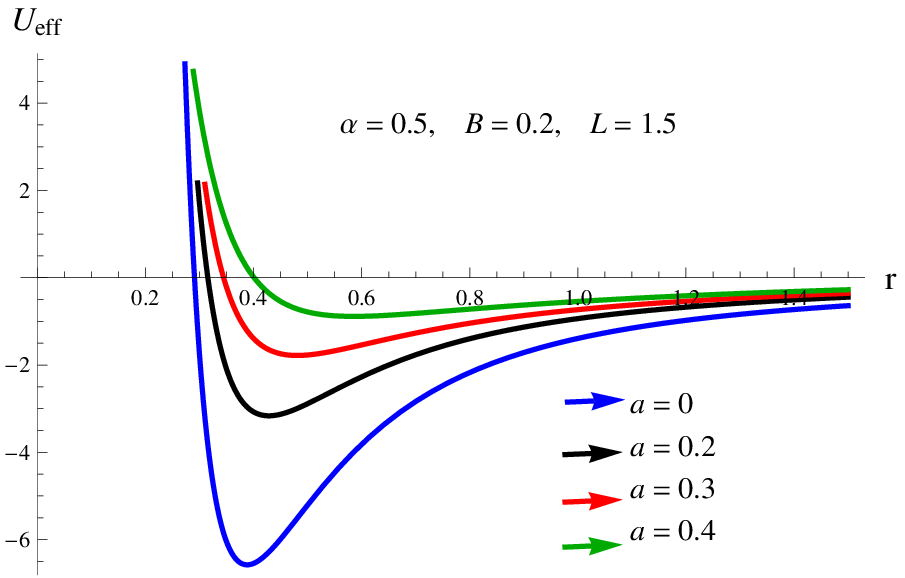,width=.44\linewidth} \caption{Effective
potential versus $r$.}
\end{figure}

\subsection{Effective Force}

The effective force acting on particle provides the information
about motion, i.e., whether it is attracted towards the BH or moving
away from it \cite{30a}. We study the particle motion in the
background of Kerr-MOG BH where attractive as well as repulsive
gravitational forces can be produced by STVG. Here, we find the
effective force acting on particle using Eq.(\ref{13}) as \cite{31}
\begin{eqnarray}\nonumber
F&=&\frac{-1}{2}\frac{dU_{eff}}{dr}\\\nonumber
&=&\frac{-GM}{2r^{4}}[L(3L+Br^{2}-6aE+24a^{2}B)
+3a^{2}E^{2}-12a^{3}BE+21a^{4}B^{2}\\\nonumber&-&B^{2}r^{4}
+r^{2}+4a^{2}B^{2}r^{2}]-\frac{(\alpha
G_{N}GM^{2})^{2}}{2r^{5}}[4BL-4aBE+B^{2}+12a^{2}B^{2}]\\\nonumber&-&\frac{(\alpha
G_{N}GM^{2})(GM)}{r^{6}}[B
L(10a^{2}-3r^{2})-B^{2}r^{4}-3aBEr^{2}+6a^{2}B^{2}r^{2}+10\\\nonumber&\times&a^{3}BE
-30a^{4}B^{2}]-\frac{(\alpha G_{N}GM^{2})^{3}}{r^{6}}40a^{4}B^{2} +
\frac{(GM)^{2}}{r^{5}}[16a^{2}B
L+48a^{4}B^{2}\\\nonumber&-&16a^{3}BE +6a^{2}B^{2}r^{2}]+
\frac{\alpha
G_{N}GM^{2}}{2r^{5}}[2L(L-2aE+6a^{2}B)+r^{2}+4a^{2}B^{2}r^{2}
\\\nonumber&-&2aBEr^{2}+2a^{2}E^{2}-12a^{3}B+18a^{4}B^{2}]+\frac{1}{2r^{3}}[L(1
+2a^{2}B)+a^{2}-3a^{4}\\\nonumber&\times&B^{2}+4a^{3}BE-a^{2}E^{2}
-B^{2}r^{4}]+\frac{(\alpha
G_{N}GM^{2})^{3}B^{2}}{r^{5}}+\frac{(\alpha
G_{N}GM^{2})^{2}(GM)}{2r^{6}}\\\nonumber&\times&[20a^{2}B^{2}-3B^{2}r^{2}]+\frac{(\alpha
G_{N}GM^{2})(GM)^{2}}{r^{7}}[27a^{4}B^{2}-16a^{2}B^{2}r^{2}].
\end{eqnarray}
We see that the first, second and third terms are attractive if
$L(3L+Br^{2}-6aE+24a^{2}B)>3a^{2}E^{2}-12a^{3}BE+21a^{4}B^{2}-B^{2}r^{4}
+r^{2}+4a^{2}B^{2}r^{2}$, $4BL>-4aBE+B^{2}+12a^{2}B^{2}$ and $B
L(10a^{2}-3r^{2})>-B^{2}r^{4}-3aBEr^{2}+6a^{2}B^{2}r^{2}+10a^{3}BE
-30a^{4}B^{2}$, respectively. The fourth term is also attractive.
The fifth, sixth and seventh terms are repulsive if
$16a^{2}BL>48a^{4}B^{2}-16a^{3}BE +6a^{2}B^{2}r^{2}$,
$2L(L-2aE+6a^{2}B)>r^{2}+4a^{2}B^{2}r^{2}
-2aBEr^{2}+2a^{2}E^{2}-12a^{3}B+18a^{4}B^{2}$ and $L(1
+2a^{2}B)>a^{2}-3a^{4}B^{2}+4a^{3}BE-a^{2}E^{2} -B^{2}r^{4}$,
respectively. The last three terms are also repulsive.

Figure \textbf{3} describes the behavior of effective force as a
function of $r$. In the upper panel, the left graph shows that the
effective force acting on particles is more attractive for large
values of magnetic field. The right graph shows the comparison of
effective force acting on a particle around Kerr-MOG BH with the
Kerr and Schwarzschild BHs. We see that the effective force on a
particle for Kerr-MOG BH attains more values than the Kerr and
Schwarzschild BHs and increases for increasing value of $\alpha$.
This means that repulsion to reach the singularity for Kerr-MOG BH
is more as compared to that in Schwarzschild and Kerr BHs. The
behavior of effective force for different values of spin parameter
is represented in lower graph showing that effective force acting on
particles increases with the increase of spin parameter and becomes
almost constant as the particle moves away from the BH. It is also
observed that effective force on particles in the vicinity of
Schwarzschild-MOG BH is small as compared to the Kerr-MOG BH.
\begin{figure}\centering
\epsfig{file=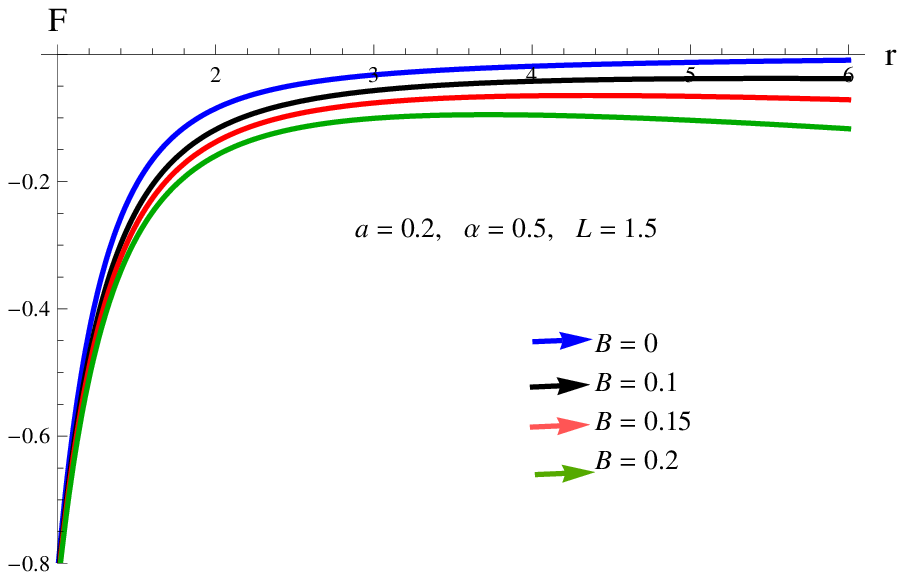,width=.44\linewidth}
\epsfig{file=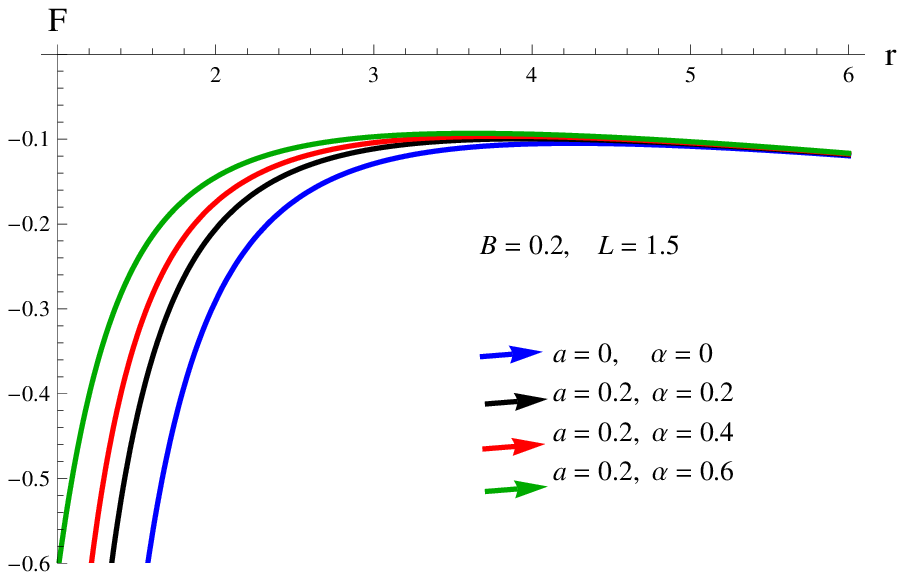,width=.44\linewidth}
\epsfig{file=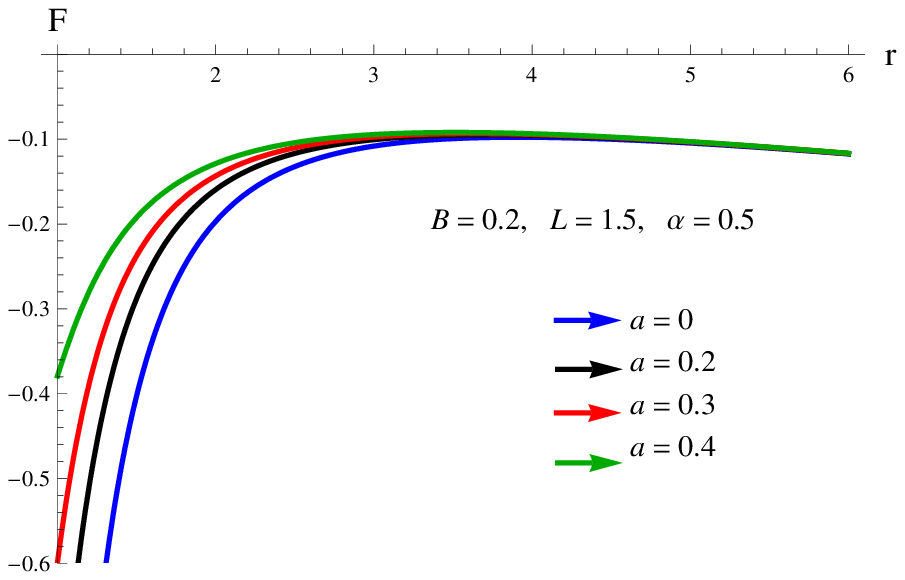,width=.44\linewidth} \caption{Effective force as
a function of $r$.}
\end{figure}

\subsection{Lyapunov Exponent}

Lyapunov exponent measures the average rate of expansion or
contraction of trajectories in a phase space. The positive and
negative Lyapunov exponents indicate divergence and convergence
between neighboring orbits \cite{32}. Using Eq.(\ref{13}), we can
find the Lyapunov exponent as \cite{33}
\begin{eqnarray}\nonumber
\lambda&=&\sqrt{\frac{-U_{eff}^{\prime\prime}(r_{0})}{2\dot{t}^{2}(r_{0})}}\\\nonumber
&=&[\frac{1}{2L^{2}r^{8}}[(r(-2GM+r)+\alpha
G_{N}GM^{2})(r(-480a^{4}B^{2}G^{3}M^{3}+160a^{2}BG^{2}\\\nonumber&\times&(3a^{2}B
-aE+L)M^{2}r-12G(7a^{4}B^{2}-4a^{3}BE-2aEL+L^{2}+a^{2}(E^{2}\\\nonumber&
+&8BL))Mr^{2}+3(-3a^{4}B^{2}+4a^{3}BE+L^{2}+a^{2}(1-E^{2}
+2B(L+6BG^{2}\\\nonumber&\times&M^{2})))r^{3}-2G(1+2B(2a^{2}B+L))Mr^{4}
+B^{2}r^{7})+\alpha
G_{N}GM^{2}(336a^{4}B^{2}\\\nonumber&\times&G^{2}M^{2}-120a^{2}B
G(3a^{2}B-aE+L)Mr+10((3a^{2}B-aE+L)^{2}-16a^{2}\\\nonumber&\times&B^{2}G^{2}M^{2})r^{2}
+24BG(2a^{2}B-a E+L)Mr^{3}
+3(1+2aB(2aB-E))r^{4}\\\nonumber&+&4B^{2}GMr^{5}+B r\alpha
G_{N}GM^{2}(60a^{2}B(2GM-r)
+20aEr-r(20L+3Br\\\nonumber&\times&(4GM+r))+10Br\alpha
G_{N}GM^{2})))]]^{\frac{1}{2}}\mid_{r=r_{0}}.
\end{eqnarray}
Figure \textbf{4} shows the graph of Lyapunov exponent as a function
of $B$. In the upper panel, the left graph indicates that Lyapunov
exponent has decreasing behavior for higher values of angular
momentum representing that orbits are more unstable for small value
of angular momentum as compared to large. The right graph gives a
comparison for Kerr-MOG BH with the Kerr, Schwarzschild and
Schwarzschild-MOG BHs. This shows that for Kerr-MOG BH, instability
of circular orbits is higher as compared to Kerr, Schwarzschild and
Schwarzschild-MOG BHs and instability increases with the increase of
$\alpha$. The behavior of Lyapunov exponent for different values of
spin parameter is shown in the lower graph. It is noted that orbits
are more unstable with large value of $a$ as compared to small.
\begin{figure}\centering
\epsfig{file=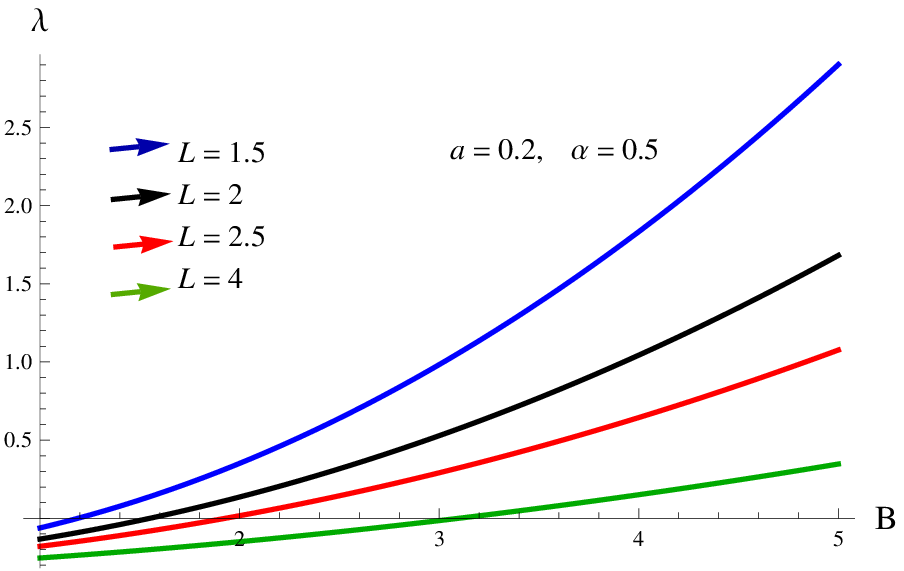,width=.44\linewidth}
\epsfig{file=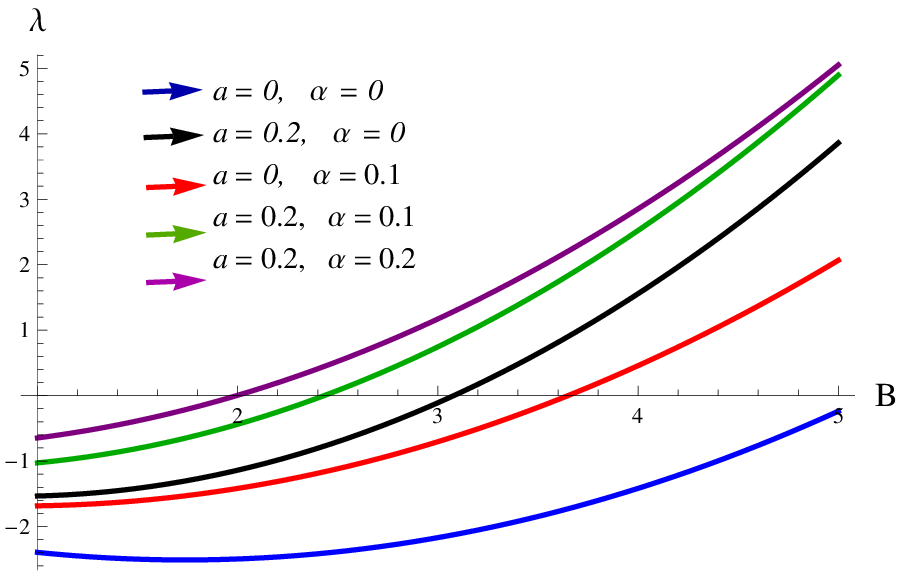,width=.44\linewidth}
\epsfig{file=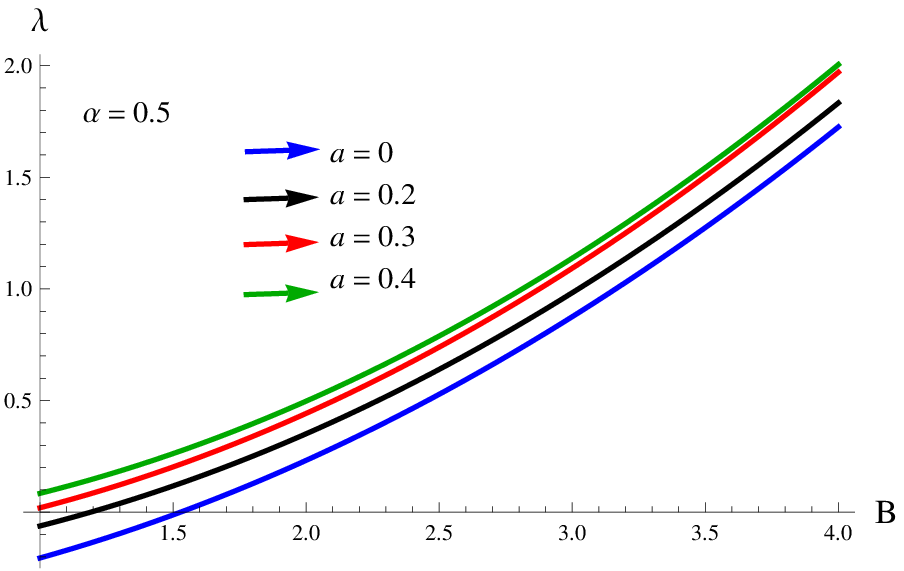,width=.44\linewidth} \caption{Lyapunov exponent
as a function of $B$ for $L=1.5$.}
\end{figure}

\section{Center of Mass Energy}

The center of mass energy of two colliding particles can be obtained
by adding their masses and kinetic energies depending upon the
interacting particles and gravitational field around the
astrophysical object. It is interesting to discuss the particle
collision as it is a naturally occurring process in the universe. In
the following, we discuss CME for neutral as well as charged
particles.

\subsection{Neutral Case}

Let us consider two neutral particles with the same rest-mass
$m_{0}$ but different four velocities $u_{1}^{\sigma}$ and
$u_{2}^{\eta}$ colliding with each other. The conserved energy and
angular momentum of colliding particles are $E_{1},~E_{2},~L_{1}$
and $L_{2}$. The CME of colliding particles is defined as
\begin{equation}\label{14}
\left(\frac{E_{cm}}{\sqrt{2}m_{0}}\right)^{2}=1-g_{\sigma\eta}u_{1}^{\sigma}u_{2}^{\eta}.
\end{equation}
Inserting the values of $g_{\sigma\eta}$, $u_{1}^{\sigma}$ and
$u_{2}^{\eta}$ from Eqs.(\ref{5})-(\ref{7}), the CME becomes
\begin{eqnarray}\nonumber
\left(\frac{E_{cm}}{\sqrt{2}m_{0}}\right)^{2}&=&\frac{1}{r^{2}\Delta}[(r(r(a^{2}+r(-2GM+r))
+(r^{3}+a^{2}(2GM+r))E_{1}E_{2}\\\nonumber&-&
2aGME_{1}L_{2}-L_{1}(2aGME_{2}+(r-2MG)L_{2})+\alpha
G_{N}GM^{2}(r^{2}\\\label{15}&-&a^{2}E_{1}E_{2}+L_{1}(aE_{2}-L_{2})+
aE_{1}L_{2}))-\sqrt{S_{1}S_{2}}],
\end{eqnarray}
where
\begin{eqnarray}\nonumber
S_{i}&=&r^{4}E_{i}^{2}+r^{2}(a^{2}E_{i}^{2}-L_{i}^{2})+(2GMr-\alpha
G_{N}GM^{2})\\\nonumber&\times&(a
E_{i}-L_{i})^{2}-\frac{\Delta}{r^{2}}, \quad i=1,2.
\end{eqnarray}
For $a=\alpha=0$ and $\alpha=0$, Eq.(\ref{15}) reduces to
Schwarzschild and Kerr BHs, respectively \cite{23}.

\subsection{Charged Case}

Here we consider particles collision in the vicinity of magnetic
field. In this case, the CME takes the following form
\begin{eqnarray}\nonumber
\left(\frac{E_{cm}}{\sqrt{2}m_{0}}\right)^{2}&=&\frac{1}{r^{4}\Delta}[-B^{2}G^{3}
M^{6}r^{2}\alpha^{3}G_{N}^{3}
+BG^{2}M^{4}r\alpha^{2}G_{N}^{2}(-8a^{2}B G
M+6a^{2}B\\\nonumber&\times& r+2BGMr^{2}
+Br^{3}-ar(E_{1}+E_{2})+r(L_{1}+L_{2}))+\alpha
G_{N}GM^{2}\\\nonumber&\times&(-16a^{4}B^{2}G^{2}M^{2}
+24a^{4}B^{2}G M r-9a^{4}B^{2}r^{2}
+16a^{2}B^{2}G^{2}M^{2}r^{2}\\\nonumber&-&8a^{2}B^{2}GMr^{3}
+r^{4}-4a^{2}B^{2}r^{4}-4B^{2}GMr^{5}+B^{2}r^{6}
+(E_{1}+E_{2})\\\nonumber&\times&(-4a^{3}B G M
r+3a^{3}Br^{2}+2aBGMr^{3}
+aBr^{4})+r(a^{2}B(4GM\\\nonumber&-&3r)-2BGMr^{2}+arE_{1})L_{2}
+rL_{1}(a^{2}B(4GM-3r)-2BGMr^{2}\\\nonumber&+&arE_{2}-rL_{2}))
-r(-32a^{4}B^{2}G^{3}M^{3}+48a^{4}B^{2}G^{2}M^{2}r
-14a^{4}B^{2}G\\\nonumber&\times&Mr^{2}-a^{2}r^{3}-3a^{4}B^{2}r^{3}
+12a^{2}B^{2}G^{2}M^{2}r^{3}+2GMr^{4}-8a^{2}B^{2}GM\\\nonumber&\times&r^{4}
-r^{5}-2a^{2}B^{2}r^{5}-2B^{2}GMr^{6}+B^{2}r^{7}
+(E_{1}+E_{2})(-8a^{3}BG^{2}M^{2}\\\nonumber&\times&
r+4a^{3}BGMr^{2} +2a^{3}Br^{3}+2aBr^{5})+E_{1}E_{2}(-a^{2}r^{2}
-a^{2}r^{3}-r^{5}\\\nonumber&-&2a^{2}GMr^{2}) +r(Br^{3}(-2GM+r)
+a^{2}B(8G^{2}M^{2}-8GMr+r^{2})\\\nonumber&+&2aGMrE_{1})L_{2}+rL_{1}(Br^{3}(-2GM+r)
+a^{2}B(8G^{2}M^{2}-8GMr\\\nonumber&+&r^{2})+2aGMrE_{2}+r(-2GM+r)L_{2})))
-\sqrt{S_{1}S_{2}}],
\end{eqnarray}
where
\begin{eqnarray}\nonumber
S_{i}&=&[r(a^{4}B^{2}(4GM-3r)(8G^{2}M^{2}-6GMr-r^{2})-4a^{3}BE_{i}
r(-4G^{2}M^{2}\\\nonumber&+&2GMr+r^{2})-4aE_{i}r^{2}(G
L_{i}M+Br^{3})+a^{2}r(-2BL_{i}(8G^{2}M^{2}
-8GMr\\\nonumber&+&r^{2})+2B^{2}r^{2}(-6G^{2}M^{2}+4GMr+r^{2})+r(-r+E_{i}^{2}(2GM+r)))
+r^{2}\\\nonumber&\times&(-L_{i}^{2}r
+(-1+E_{i}^{2}-2BL_{i})r^{3}-B^{2}r^{5}+2GM(r^{2}+
(L_{i}+Br^{2})^{2})))\\\nonumber&-&\alpha
G_{N}GM^{2}(16a^{4}B^{2}G^{2}M^{2}-8a^{2}B
G(3a^{2}B-aE_{i}+L_{i})Mr+((3a^{2}B\\\nonumber&-&aE_{i}+L_{i})^{2}
-16a^{2}B^{2}G^{2}M^{2})r^{2}+4BG(2a^{2}B-aE_{i}+L_{i})Mr^{3}
+(1\\\nonumber&+&2aB(2aB-E_{i}))r^{4}+4B^{2}GMr^{5}-B^{2}r^{6}+Br\alpha
G_{N}GM^{2}(a^{2}B(8GM\\\nonumber&-&6r)+2aE_{i}r-r(2L_{i}+B
r(2GM+r))+Br\alpha G_{N}GM^{2}))].
\end{eqnarray}
Clearly, the CME will be infinite if one of the colliding particles
have diverging angular momentum at horizon. Thus, for finite CME
only finite values of angular momentum are allowed. The behavior of
CME as a function of $r$ in the absence as well presence of magnetic
field is depicted in Figure \textbf{5}. In the upper panel, the left
graph gives a comparison of CME for colliding particle near Kerr-MOG
with Schwarzschild and Kerr BH. The collision occurring near the
horizon of Kerr-MOG BH can produce high energy as compared to
Schwarzschild and Kerr BHs and increases with the increase of
parameter $\alpha$. The right graph is plotted for different values
of spin parameter. We see that CME strongly depends on the rotation
of BH. High energy can be achieved with the maximum spin. The CME
for different values of parameter $\alpha$ (left) and spin parameter
(right) in the presence of magnetic field is shown in the lower
panel. It is noted that maximum CME can be produced in the presence
of magnetic field as compared to its absence. The CME has decreasing
behavior with the increase of radial distance $r$ and becomes almost
constant away from the event horizon of BH. The CME for colliding
particles does not diverge in the absence/presence of magnetic
field.

Figure $\textbf{6}$ is depicted for different values of $L_{1}$ in
the presence as well absence of magnetic field. We see that the CME
decreases with the increase of angular momentum. The particle
colliding with small angular momentum can produce high energy as
compared to particle with large angular momentum. Initially, CME
decreases and then becomes constant with the increase of radial
distance $r$. It is also observed that CME is finite for finite
values of angular momentum and attain more energy in the presence of
magnetic field as compared to absence.
\begin{figure}\centering
\epsfig{file=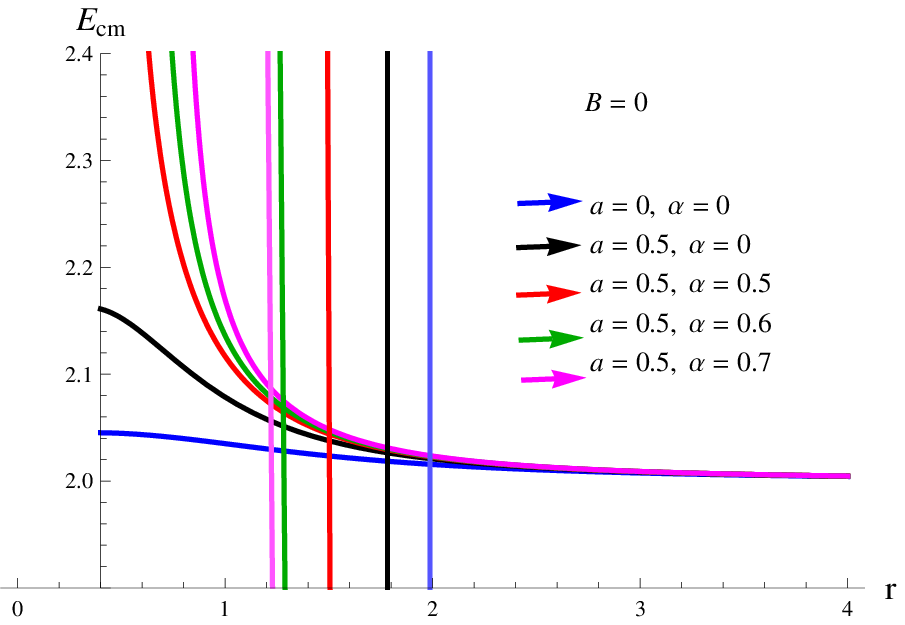,width=.44\linewidth}
\epsfig{file=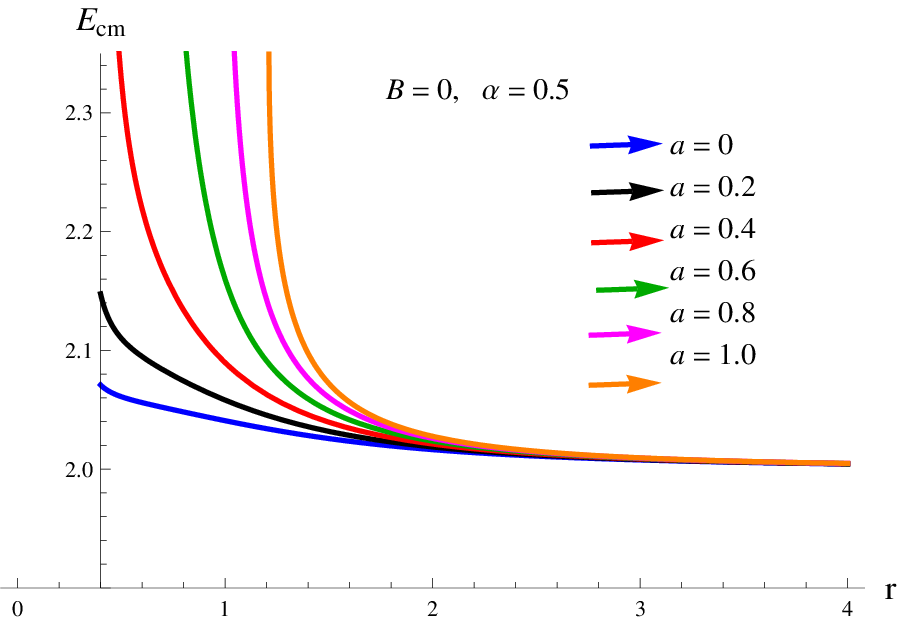,width=.44\linewidth}
\epsfig{file=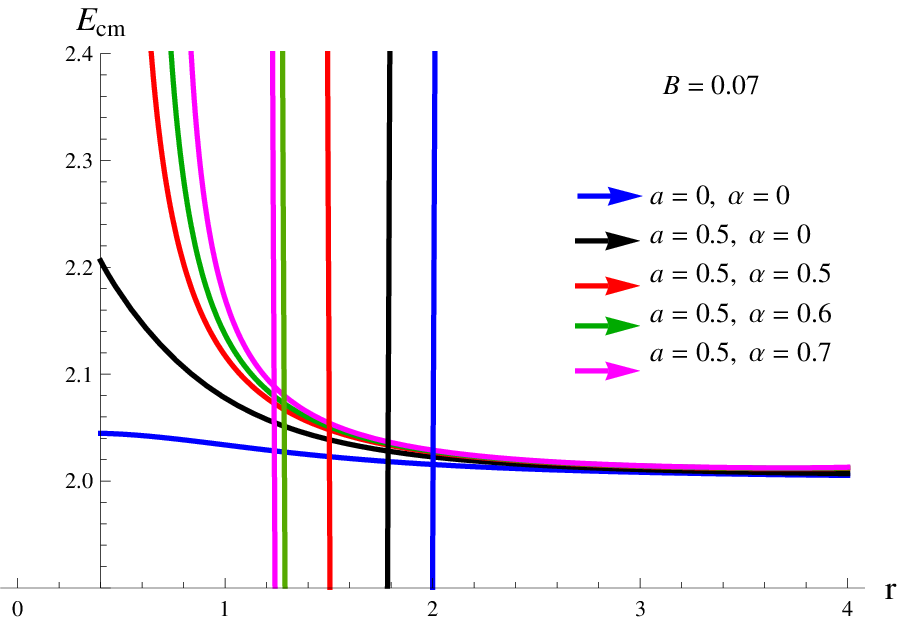,width=.44\linewidth}
\epsfig{file=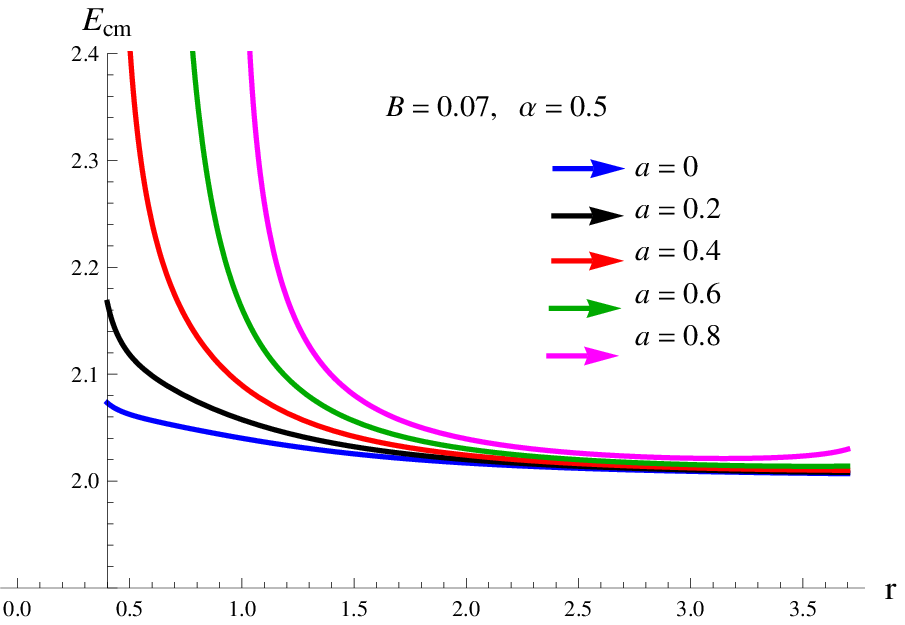,width=.44\linewidth} \caption{The CME with
respect to $r$ for $L_{1}=1$ and $L_{2}=1.5$. Here, the vertical
lines are event horizons.}
\end{figure}
\begin{figure}\centering
\epsfig{file=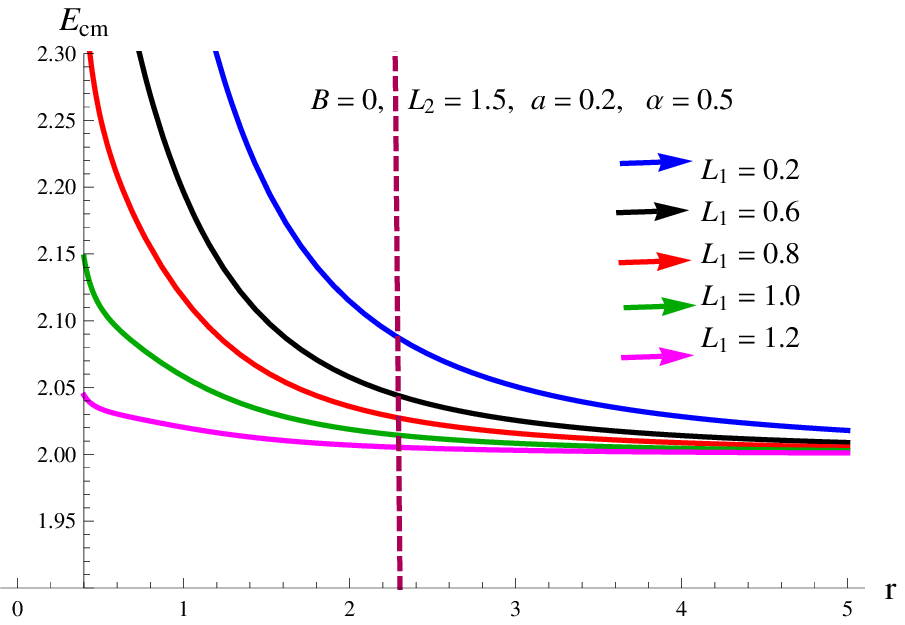,width=.44\linewidth}
\epsfig{file=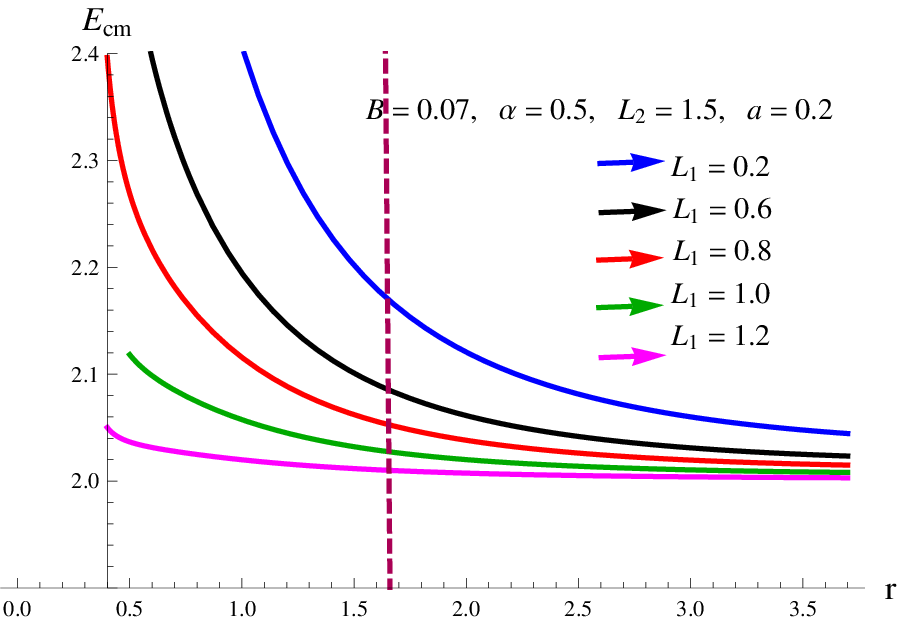,width=.44\linewidth} \caption{The CME as a
function of $r$. The vertical line represent the event horizon.}
\end{figure}

\section{Concluding Remarks}

In this paper, we have studied the dynamics of particles around the
Kerr-MOG BH in the absence/presence of magnetic field. We have
explored geodesics for both neutral as well as charged particles. We
have graphically discussed conditions for a particle to escape to
infinity after its collision with another particle. The effect of
magnetic field, angular momentum, parameter $\alpha$ as well as spin
parameter $a$ on the motion of neutral and charged particles is also
analyzed graphically. It is seen that the escape velocity increases
with the increase of $\alpha$ and magnetic field but decreases with
the increase of $L$. The particles can attain more energy in the
presence of magnetic field and can escape easily. The escape
velocity also depends upon the spin of BH. There will be more
possibilities to escape to infinity for large value of spin
parameter. It is found that the escape velocity of particle around
Schwarzschild BH is smaller as compared to Kerr and Kerr-MOG BH.
Particles cannot escape easily in the vicinity of Kerr-MOG BH as
compared to Kerr, Schwarzschild and Schwarzschild-MOG BHs.

We have explored stability of circular orbits by the effective
potential. It is observed that the effective potential increases for
large values of magnetic field which indicates that the presence of
magnetic field increases the stability of particles orbits. We have
compared the stability of circular orbits around Kerr-MOG BH with
the Kerr and Schwarzschild BHs which indicates that circular orbits
around Kerr-MOG BH are more unstable as compared to Kerr,
Schwarzschild and Schwarzschild-MOG BHs. We note that large magnetic
field leads to unstable motion. The rotation of BH have strong
effects on the stability of orbits which decreases with the increase
of $a$. We have also discussed the instability of circular orbits
through Lyapunov exponent as a function of magnetic field $B$.

Finally, we have calculated the CME for two interacting particles
around the Kerr-MOG BH. It is found that particle collision can
produce high energy near Kerr-MOG BH as compared to Kerr,
Schwarzschild and Schwarzschil-MOG BHs. We observe that the CME
increases with the increase of spin parameter as well as parameter
$\alpha$ but decreases with the increase of angular momentum. The
CME is finite for finite values of angular momentum and can be
infinite for diverging angular momentum. We conclude that the
external magnetic field, parameter $\alpha$, spin parameter affect
the motion of particles in STVG. It is worth mentioning here that,
our work is the generalization of \cite{12} reduce to
Schwarzschild-MOG BH \cite{12} when $a=0$ and to Schwarzschild BH
\cite{34} for $a=\alpha=0$.

\end{document}